\documentclass[aps,prc,twocolumn,superscriptaddress,showpacs,nofootinbib]{revtex4-1}
\usepackage{graphicx,epstopdf,amsmath,amssymb,bm,color}

\newcommand{\beq}{\begin{equation}}
\newcommand{\eeq}{\end{equation}}

\allowdisplaybreaks

\begin{document}

\title{Diffusion Monte Carlo study of strongly interacting two-dimensional Fermi gases}

\author{Alexander Galea}
\affiliation{Department of Physics, University of Guelph, Guelph, Ontario N1G 2W1, Canada}
\author{Hillary Dawkins}
\email[Present address:~]{Department of Physics and Astronomy, University of Waterloo, Waterloo, Ontario N2L 3G1, Canada}
\affiliation{Department of Physics, University of Guelph, Guelph, Ontario N1G 2W1, Canada}
\author{Stefano Gandolfi}
\affiliation{Theoretical Division, Los Alamos National Laboratory,
Los Alamos, New Mexico 87545, USA}
\author{Alexandros Gezerlis}
\affiliation{Department of Physics,
University of Guelph, Guelph, Ontario N1G 2W1, Canada}

\begin{abstract}
Ultracold atomic Fermi gases have been a popular topic of research, with attention being
paid recently to two-dimensional (2D) gases. In this work, we perform $T=0$ {\it ab initio} diffusion Monte Carlo calculations for a strongly interacting two-component Fermi gas confined to two dimensions.  We first go over finite-size systems and the connection to the 
thermodynamic limit. After that, we 
illustrate pertinent 2D scattering physics and properties of the 
wave function. We then show energy results for the strong-coupling crossover, 
in between the Bose-Einstein Condensation (BEC) and Bardeen-Cooper-Schrieffer (BCS)
regimes.  
Our energy results for the BEC-BCS crossover are parametrized to produce an equation of state, which is used to determine Tan's contact.  We carry out a detailed comparison with other 
microscopic results.  Finally, we calculate the pairing gap for a range of interaction strengths in the strong coupling regime, following from variationally optimized many-body wave functions.
\end{abstract}


\maketitle

\section{Introduction}

Cold atomic gases have seen a number of experimental 
breakthroughs, which have allowed detailed testing of theoretical models~\cite{Bloch:2008,Giorgini:2008}.  The strongly paired Fermi gas is accessible through the use of 
Feshbach resonances, where an external magnetic field is used to control interactions between cold atoms.  This has allowed for measurements of several gas properties 
in the BEC-BCS crossover.  These experiments involve two-component gases of spin-up and spin-down particles interacting at zero-range.  In the heart of the crossover, for 3D gases, a unitary regime exists where the dilute gas behaves universally (is scale independent).  Quantum Monte Carlo (QMC) methods can be used to calculate ground-state properties for any interaction strength in the crossover.  Such {\it ab initio} methods, solving the problem from first principles, provided the first quantitatively reliable prediction of properties of the unitary gas~\cite{Carlson:2003}.

A duality is expected between superfluid gases of Fermi atoms and neutrons in compact stars. 
Neutron-matter properties depend on the interaction strength (roughly, a product of the density and scattering length). This means that we can learn about neutron pairing by studying the Fermi gas on the BCS side of the crossover~\cite{Gezerlis:2008,Carlson:2012,Gandolfi:2015}. 
Unlike atoms, the neutron-neutron interaction cannot be tuned microscopically.  Instead, the density impacts the pairing properties. Furthermore, the 
neutron-neutron effective range is finite, comparable in size to the average interparticle spacing. 

The pairing properties of two-dimensional quantum gases are being intensely studied more recently.  Part of this interest is generated by the possibility of connections to
condensed-matter research (e.g., high-temperature superconductors and Dirac fermions in graphene).  In neutron stars, nuclear ``pasta'' phases are expected, where neutrons are restricted to 1D or 2D configurations. More generally, reducing the 
dimensionality can lead to new phenomena as well as new insights on many-body quantum 
theory.  A detailed theoretical study is further justified by 2D Fermi gas experiments~\cite{Gunter:2005,Martiyanov:2010,Frohlich:2011,Feld:2011,Orel:2011,Makhalov:2014,Ong:2015,Murthy:2015,Ries:2015,Fenech:2015,Boettcher:2015}. In these studies, the 3D quantum degenerate gas can be separated into a series of pancake shaped, quasi-2D gas clouds using a highly anisotropic trapping potential.

The BCS theory was first applied to 2D Fermi gases in the 1980's~\cite{Miyake:1983,Randeria:1989}. These mean-field calculations
can qualitatively describe either large or small pair sizes, corresponding to the BCS or BEC limits, respectively.  The intermediate regime is not expected to be accurately described by a mean-field theory.  A 2011 QMC study estimated the ground state energy and pairing gap for a range of interaction strengths~\cite{Bertaina:2011}.  This was the first {\it ab initio} equation of state (EOS) prediction for the strongly interacting 2D Fermi gas.  More recently, other
QMC methods have been used to tackle this problem, both at zero and at finite temperature~\cite{Shi:2015,Anderson:2015}. The special appeal of non-perturbative methods is their 
ability to provide dependable calculations of quantities in the center of the BEC-BCS crossover.
In parallel, other many-body approaches have also emerged, leading to equations of state for 
the 2D gas
that are qualitatively similar to the QMC values~\cite{Bauer:2014,He:2015,Klawunn:2015}.

In this work, we study the 2D two-component $T=0$ Fermi gas using diffusion Monte Carlo (DMC).  We pay close attention to the problem of variational minimization and the conclusions 
that can be drawn from it. This work is organized as follows.  First we discuss the noninteracting Fermi gas (Sec.~\ref{sec:free}), with emphasis on finite-size effects associated with periodic simulation areas.  Interactions are the focus of Sec.~\ref{sec:scatter}, where 2D s-wave scattering theory for low-energy collisions is illustrated.  We then combine topics from the previous two sections and formulate the many-body problem with interactions (Sec.~\ref{sec:interacting}).  The QMC methods are outlined in Sec.~\ref{sec:QMC}, before moving on to results.  Our ground-state energy and contact parameter results are shown and discussed in Sec.~\ref{sec:results}, where we also compare to previous QMC studies.  Finally, the pairing gap calculation is described and our results are shown in Sec.~\ref{sec:gap}.  We conclude with a discussion of related future research avenues.

\section{2D non-interacting Fermi gas}
\label{sec:free}

A distinguishing feature of the zero-temperature Fermi gas is the occupation of finite-momentum levels due to the Pauli exclusion principle.  We are interested in solving the interacting many-body problem, but it's useful to first consider the simple system of free particles.  In this section we will focus on the energy of finite-size 2D systems.

Take $N$ particles in a uniform box of length $L$ with periodic boundary conditions.  The thermodynamic-limit (TL) gas is a limiting case of this system where particle number $N\rightarrow \infty$ and simulation area $A\rightarrow \infty$, while $N/A \rightarrow {\rm constant}$.  Single particle states are written as plane waves: $\psi_{\bf k}({\bf r}) = e^{i{\bf k_n} \cdot {\bf r}} / L$, where momentum levels are identified by the wave vector:
\begin{equation}
\label{eq:wavevector}
{\bf k_n} = \frac{2\pi}{L} (n_x\hat{x}+n_y\hat{y})\,.
\end{equation}
Each particle has energy $E_{\bf n} = \hbar^2 {\bf k_n}^2 /2m$, where $m$ is the mass and the energy level is identified by ${\bf n}=(n_x,n_y)$.  Identical half-odd spin particles must be in distinct states and, at $T=0$, will occupy all available levels up to the Fermi surface.  Particles on the Fermi surface have energy $\epsilon_F = \hbar^2 k_F^2 /2m$, where $k_F$ is the Fermi wave number.

\begin{figure}[t]
\centering
\includegraphics[width=0.45\textwidth]{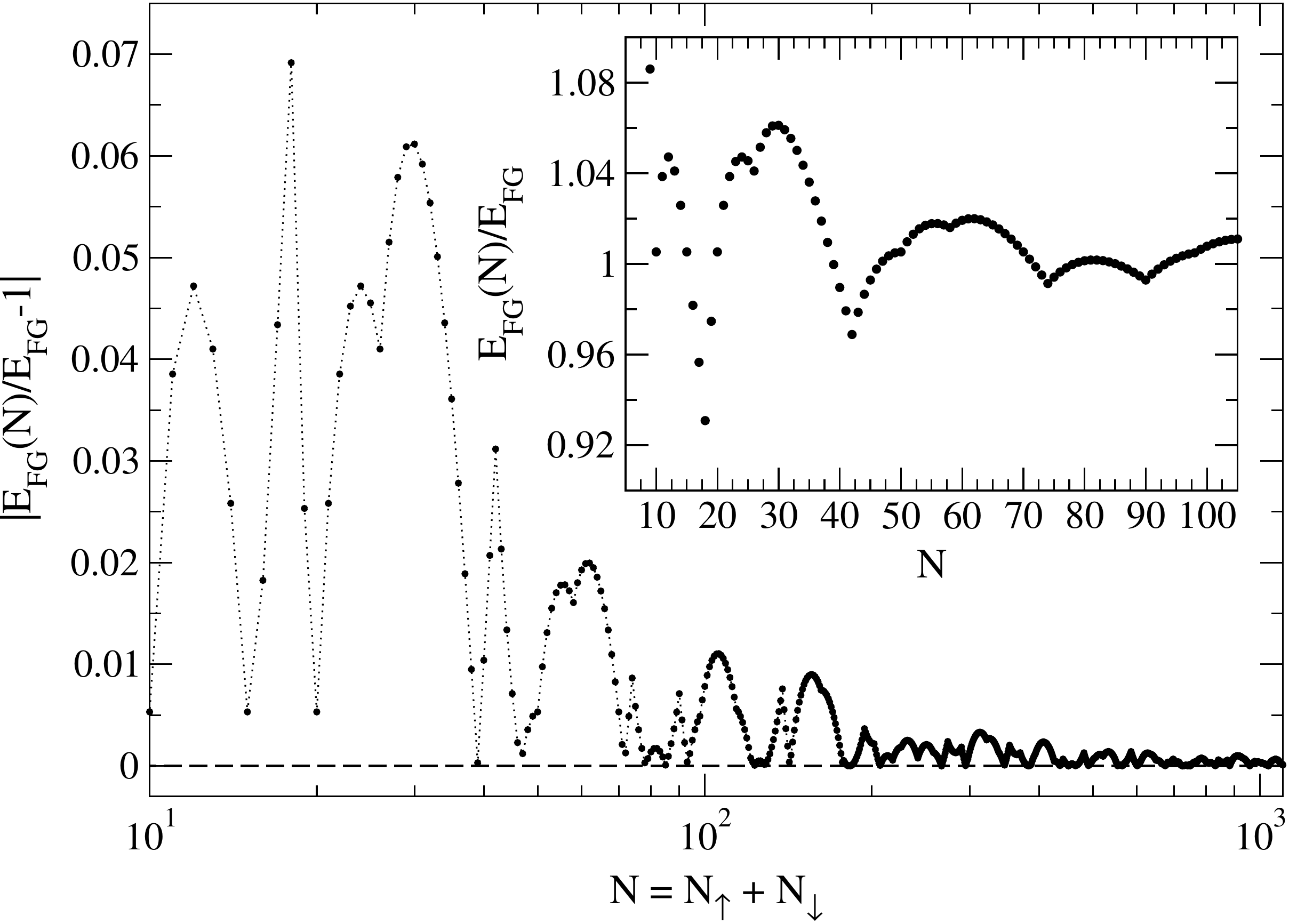}
\caption{Error associated with finite-size periodic systems, defined in units of $E_{\rm FG}$ as the absolute value of the energy difference between $N$ particles in a finite size box and the TL system where $N\rightarrow \infty$ and $A\rightarrow \infty$.  The largest error is seen for small $N$ where dotted lines are drawn to guide the eye.  The inset displays this range on a linear scale.  For large values of $N$ the error approaches zero as marked by the dashed line. \label{fig:finite}}
\end{figure}

In the TL, energy per particle for the free gas is $E_{\rm FG} = \epsilon_F /2$, and $k_F = \sqrt{2\pi n}$.  As a reminder, the corresponding expressions in 3D are $E_{\rm FG} = (3/5)\epsilon_F$ and $k_F = (3\pi^2 n)^{1/3}$.  The energy per particle for finite size systems, $E_{\rm FG}(N)$, has the largest differences from $E_{\rm FG}$ for small $N$.  This relative error is shown in Fig.~\ref{fig:finite} and gradually tends towards zero as $N$ increases.  When $N$ is even, the system has an equal number of spin-up and spin-down particles.  Points with odd $N$ are determined by placing either a spin-up or spin-down particle into the next available momentum state.  The inset shows how $E_{\rm FG}(N)$ fluctuates above and below the TL result at small $N$.

Different combinations of $n_x$ and $n_y$ in Eq.~(\ref{eq:wavevector}) can result in the same $|{\bf k_n}|$, meaning particles in different states may have the same energy. Closed-shell configurations exist at specific values of $N$ for finite-sized systems where the population in each energy-level is maximal.  By considering equal numbers of spin-up and spin-down particles for even $N$, the population allowed in each energy level doubles compared to a one-component
gas. In 2D, closed shells occur at total particle number $N$=\{2, 10, 18, 26, 42, 50, 58\dots\}, which roughly correspond to the local minima in the inset of Fig.~\ref{fig:finite}. 
For comparison, we note that in 3D the corresponding closed shells occur at 
total particle number $N$=\{2, 14, 38, 54, 66, 114, 162\dots\}.

\section{two-body scattering}
\label{sec:scatter}
Scattering in 2D for a finite-range potential $V(r)$ is described by the Schr\"odinger equation. The full wave function can be separated into radial and angular functions: $\psi(r,\theta) = R(r)\,T(\theta)$. Focusing on s-wave scattering implies setting the orbital angular momentum 
$l=0$, so the angular part must satisfy the relationship $\partial^2 T(\theta) / \partial \theta^2 = 0$.  To simplify the full 2-body Schr\"odinger equation, we define $u(r) = \sqrt{r}R(r)$ as the 2D reduced radial wave function and find:
\begin{equation}
\label{eq:swave}
-\frac{\partial^2 u(r)}{\partial r^2} = u(r) \bigg[k^2 - \frac{2m_r}{\hbar^2} V(r) + \frac{1}{4r^2}\bigg] \,,
\end{equation}
where $m_r$ is the reduced mass of the interacting bodies and $k^2$ is proportional to the scattering energy.  This equation can be solved numerically, 
carefully selecting the boundary conditions given 
the presence of the singularity at $r=0$.

Unlike the 3D case, purely attractive potentials in 2D support bound states for any strength.  Qualitatively, by confining particles to scatter in a plane, it becomes impossible for bodies with pairwise attractive interactions to escape each other.  
This is related to the $1/r^2$ dependence in Eq.~(\ref{eq:swave}) that persists even for purely s-wave scattering where there is no 
$-l^2/r^2$ centrifugal barrier.  At small $r$ this term becomes large and dramatically affects wave function solutions, despite having an increasingly small contribution as $r \rightarrow \infty$.  On the other hand, in 3D the reduced radial wave function is defined as $u_{\rm 3D}(r) = r R(r)$ and the Schr\"odinger equation has no $1/r^2$ dependence for s-wave scattering.  
As an aside, the fact that in 2D a bound state always forms regardless of the strength of the
attractive interaction brings to mind the Cooper-pair problem, though the latter is
a many-body (i.e., beyond two-body) effect.

\begin{figure}[b]
\centering
\includegraphics[width=0.45\textwidth]{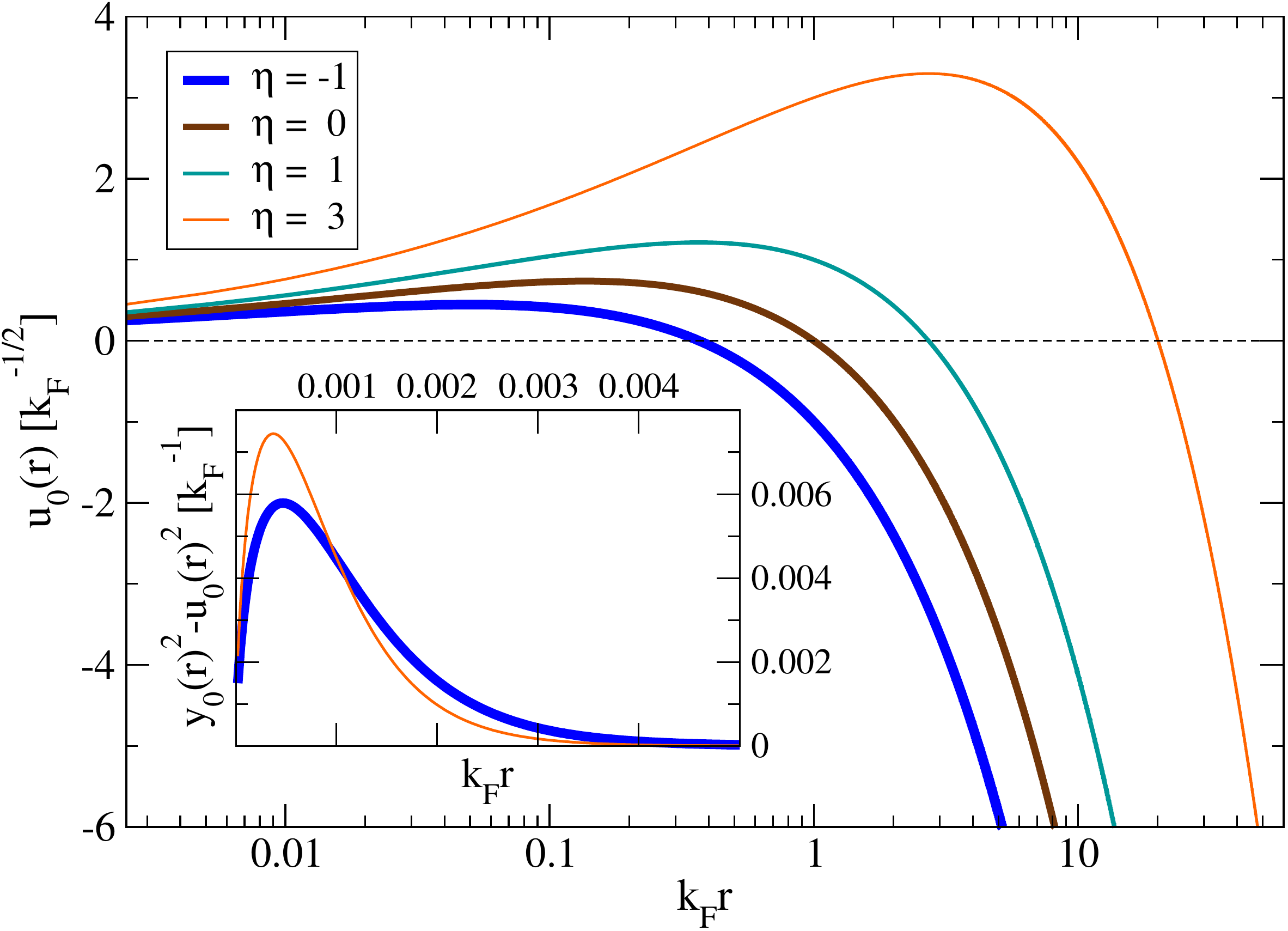}
\caption{Wave function solutions to Eq.~(\ref{eq:swave}) in the limit of zero scattering energy ($k^2 \rightarrow 0$) for the modified P\"oschl-Teller potential.  Parameters $\mu$ and $v_0$ in $V(r)$ are varied to describe a range of interaction strengths $\eta = \ln(k_F a_{\rm 2D})$, labeled by line thickness.  The scattering length is defined as the point where the wave function crosses the $r$-axis, which is marked with a dashed line.  The inset shows the effective range integrand in Eq.~(\ref{eq:re2d}) for two interaction strengths, $\eta=-1$ (thick line) and $\eta=3$ (thin line). \label{fig:2dscatter}}
\end{figure}

Regardless of the detailed features of $V(r)$, scattering can be characterized by the scattering length $a_{\rm 2D}$ and effective range $r_{e}$.  These scattering parameters reflect how $u(r)$ is affected by the potential at low scattering energies.  In order to determine them we define $u_0(r)$ as the solution to Eq.~(\ref{eq:swave}) in the limit $k\rightarrow 0$ and compare with its asymptotic form $y_0$ defined by $\partial^2 y_0(r) / \partial r^2 = - y_0(r)/4r^2$.  We use the solution:
\begin{equation}
\label{eq:y0}
y_0(r) = - \sqrt{r} \ln(r/a_{\rm 2D})\,,
\end{equation}
and normalize $u_0(r)$ to match $y_0(r)$ outside the range of the potential.  $a_{\rm 2D}$ is simply the $r$-intercept of $u_0(r)$ (as seen in Fig.~\ref{fig:2dscatter}, further discussed below), provided the potential is of sufficiently
short range. This is analogous to the interpretation of the $r$-intercept of $y_0(r)$ in the 3D problem.

The effective range is related to the area between $u_0(r)$ and $y_0(r)$ as defined by the integral~\cite{Adhikari:1986}:
\begin{equation}
\label{eq:re2d}
r_e^2 = 4 \int_{0}^{\infty}(y_0^2 - u_0^2)\,dr \,,
\end{equation}
and is the second-order term in the effective-range expansion relating low-energy phase shifts $\delta(k)$ to the scattering parameters $a_{\rm 2D}$ and $r_e$.  In 2D for small values of $k$~\cite{Khuri:2009}:
\begin{equation}
\label{eq:shape}
\cot \delta(k) \approx \frac{2}{\pi} [\gamma + \ln(\frac{ka_{\rm 2D}}{2})] + \frac{k^2 r_e^2}{4} \,,
\end{equation}
where $\gamma\approx 0.577215$ is Euler's constant.  Another name for this equation is the shape-independent approximation.  Any well selected function can be tuned to reproduce the desired $a_{\rm 2D}$ and $r_e$, therefore low energy scattering can be described by a broad range of potentials.  The logarithmic scattering-length dependence in Eq.~(\ref{eq:shape}) is characteristic of 2D interactions; a logarithmic dependence also appears in the asymptotic solution $y_0(r)$, Eq.~(\ref{eq:y0}).

\begin{figure}[t]
\centering
\includegraphics[width=0.45\textwidth]{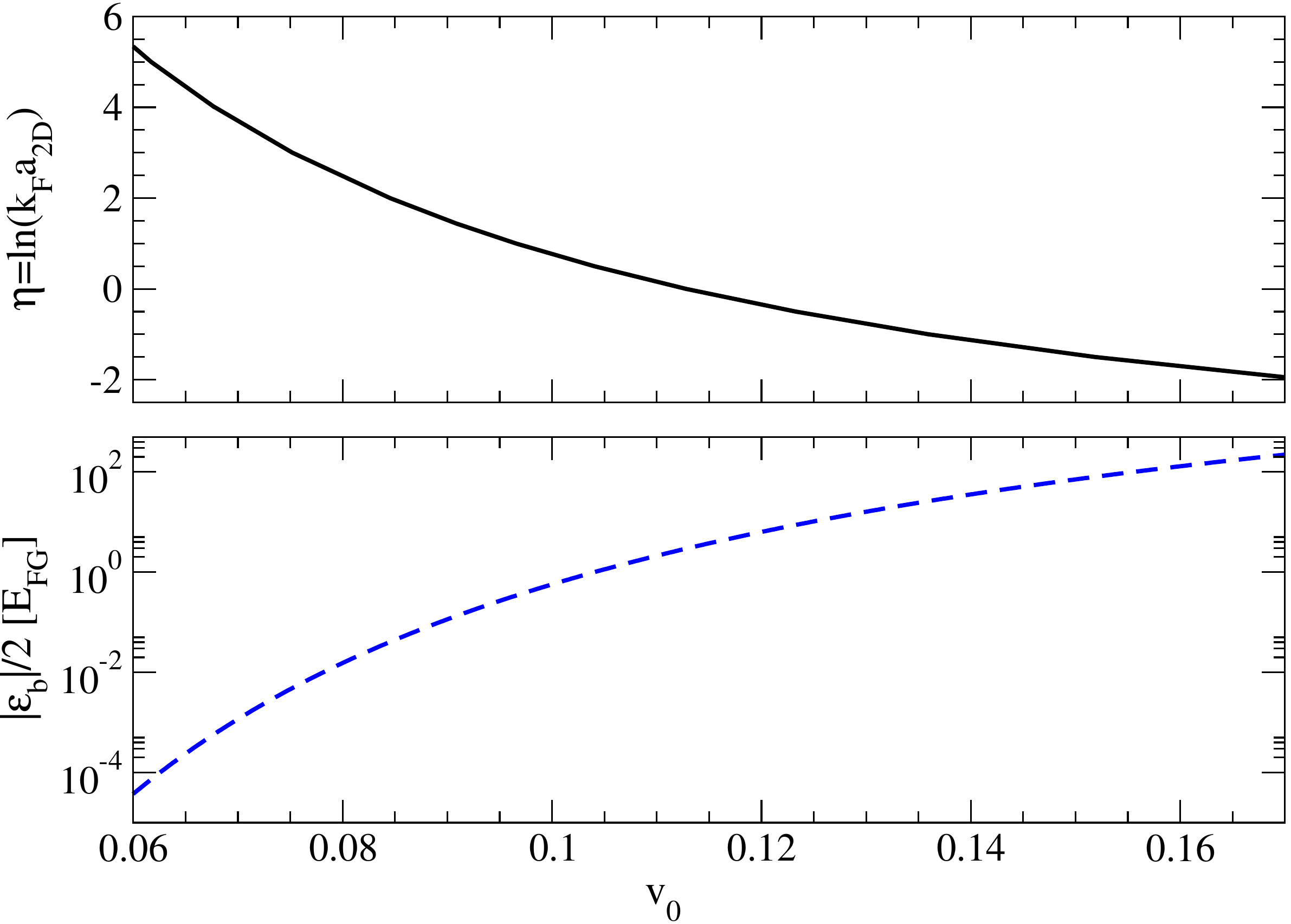}
\caption{The relationship between interaction strength $\eta = \ln(k_F a_{\rm 2D})$ and $v_0$ of the modified P\"oschl-Teller potential $V(r)$ is shown in the top panel.  The effective range is maintained constant at $k_Fr_e=0.006$ by tuning $\mu$ and $v_0$ in unison.  The absolute value of the binding energy per particle $\epsilon_b/2$ is plotted in the bottom panel.  A dramatic increase in $|\epsilon_b|$ is seen as paired particles form increasingly stronger bound states. \label{fig:v0log_eb}}
\end{figure}

We note that a difference exists for the formation of bound states in 2 and 3 dimensions.  As the depth of $V(r)$ is increased, the scattering length $a_{\rm 2D}$ approaches $0$ then diverges to $+\infty$ when a new bound state is created; in contradistinction to this, 
the 3D scattering length changes from $-\infty$ to $+\infty$ when a new bound state is formed.  In both cases, scattering length of $+\infty$ corresponds to a weakly bound state that becomes tighter as the scattering length decreases.  The particle pair in 2D has binding energy given by $\epsilon_b = - 4\hbar^2 / (m a^2_{\rm 2D} e^{2\gamma})$.

In this work, we use the fully attractive modified P\"oschl-Teller potential:
\begin{equation}
\label{eq:potential}
V(r) = - v_0\, \frac{\hbar^2}{m_r}\, \frac{\mu^2}{\cosh^2(\mu r)} \,.
\end{equation}
The parameters $v_0$ and $\mu$ roughly correspond to the depth and inverse width, respectively, and are tuned such that $V(r)$ reproduces desired scattering parameters $a_{\rm 2D}$ and $r_e$.  To probe the strong-coupling crossover, calculations are done for a range of interaction strengths $\eta = \ln(k_F a_{\rm 2D})$.  The density of the many-body system (below) 
is fixed such that $k_F$ is constant, and we vary $a_{\rm 2D}$ instead.  The diluteness requirement is satisfied by taking $r_e \ll r_0$ where $r_0=1 / \sqrt{\pi n}$ is the mean interparticle spacing.  We maintain a constant effective range of $k_Fr_e=0.006$~\cite{Note:1} by adjusting $\mu$ as $v_0$ is varied.  The resulting dependence of interaction strength on $v_0$ is shown in the top panel of Fig.~\ref{fig:v0log_eb}.  We are interested in values of $v_0$ where only one bound state is supported.  Corresponding binding energies are plotted in the bottom panel for a fixed value of $\mu/k_F \approx 900$.  As $|\epsilon_b|$ becomes increasingly small (on the left side of this figure), the BCS limit is approached; the large $|\epsilon_b|$ limit corresponds to a BEC gas of composite fermion pairs.

Figure~\ref{fig:2dscatter} displays the wave function $u_0(r)$ plotted for various interaction strengths.  The logarithmic behaviour of the asymptotic form, Eq.~(\ref{eq:y0}), is illustrated.  Differences near the origin between $u_0(r)$ and $y_0(r)$ are indistinguishable on the main plot because we tune $V(r)$ to be extremely short range.  These differences are shown in the inset where the integrand in Eq.~(\ref{eq:re2d}) is plotted for a couple interaction strengths.  We are maintaining constant $r_e$, therefore the areas under each curve are equal.

Finally, we note the general differences between the BEC-BCS crossover in 2D and 3D.  As discussed and illustrated, the scattering length is positive in 2D for all $\eta$.  Because of this, 
the identification of the crossover point is more \textit{ad hoc} (see, however, the 
dotted line in Fig.~\ref{fig:gap}).
The BCS limit corresponds to $k_F a_{\rm 2D} \gg 1$ and the BEC limit to $k_F a_{\rm 2D} \ll 1$.  In 3D, a bound state does not exist for arbitrarily weak interaction, so the scattering length can be negative.  For the strongly interacting 3D gas, where the interaction strength is $\eta_{\rm \,3D} = k_F a_{\rm 3D}$, the crossover occurs at the unitary point when $\eta_{\rm \,3D} \rightarrow \pm \infty$.  If plotted in the same style as the top panel of Fig.~\ref{fig:v0log_eb}, this would correspond to a divergence of the interaction strength at some potential depth $v_0$.  To the left, on the BCS side of the crossover, the scattering length would be negative and, in the weakly paired limit, $\eta_{\rm \,3D} \rightarrow 0$ from below.  On the BEC side of the crossover, a bound state exists and the scattering length is positive.  This regime would exist to the right of the unitary point and, in the strongly paired limit, $\eta_{\rm \,3D} \rightarrow 0$ from above.

\section{Interacting many-body problem}
\label{sec:interacting}
We consider a two-component Fermi gas described by the Hamiltonian
\begin{gather}
\hat{H} = \frac{-\hbar^2}{2m} \Bigg[\sum_{i=1}^{N_\uparrow}\nabla_i^2
+\sum_{j'=1}^{N_\downarrow}\nabla_{j'}^2\Bigg]
+\sum_{i,j'}V(r_{ij'})\,,
\end{gather}
where $m_\uparrow = m_\downarrow=m$ is the particle mass and the total particle number is given by $N_\uparrow + N_\downarrow = N$.  As described in section~\ref{sec:scatter}, we take $V(r_{ij'})$ to be of the modified P\"oschl-Teller form, Eq.~(\ref{eq:potential}).  We consider
 s-wave interactions between opposite-spin particles.

The wave function for a collection of free particles can be constructed by taking an anti-symmetrized combination of single-particle plane-wave states:
\begin{equation}
\label{eq:Slater}
\Phi_{\rm S}({\bf R}) = {\cal A}\prod_{i=1}^{N_\uparrow}\psi_{{\bf k}\uparrow}({\bf r}_i) \cdot {\cal A}\prod_{j'=1}^{N_\downarrow}\psi_{{\bf k}\downarrow}({\bf r}_{j'})\,,
\end{equation}
where ${\bf R} = {\bf r}_1, \dots, {\bf r}_N$ is the many-body configuration vector labeling the position of each particle, and ${\cal A}$ is the anti-symmetrizing operator.  This standard fermionic Slater wave function can be expressed as a determinant and is therefore convenient for computational methods.  It is effective at describing the normal state of Fermi liquids where pairing is absent or weak and can be sufficiently accounted for with the introduction of Jastrow correlations.  The Jastrow-Slater many-body trial wave function is given by:
\begin{equation}
\Psi_T({\bf R}) = \prod_{ij'}f_{J}(r_{ij'}) \, \Phi_{\rm S}~,
\end{equation}
where the Jastrow terms, $f_{J}(r_{ij'})$, include the short-range correlations between interacting particles.

When pairing is strong the Jastrow-Slater wave function is inadequate.  In this work, to describe the strongly interacting Fermi gas, we use the Jastrow-BCS many-body trial wave function:
\begin{equation}
\begin{split}
\Psi_T({\bf R}) = \prod_{ij'}f_{J}(r_{ij'}) \, \Phi_{\rm BCS} \,,~~~~~ \\
\label{eq:BCS}
\Phi_{\rm BCS} = {{\cal A}} [\phi({\bf r}_{11'}) \phi({\bf r}_{22'}) ... \phi({\bf r}_{N_\uparrow N'_\downarrow})]\,.
\end{split}
\end{equation}
where we assumed $N_\uparrow = N_\downarrow$.
As in the Slater case, the anti-symmetry requirement of $\Psi_T$ for the Fermi gas is enforced by the operator $\cal A$.  Evaluating this results in a determinant of pairing functions for all interacting particles.  This ansatz has been used before in 3D Fermi gas systems~\cite{Chang:2004,Astrakharchik:2004,Gandolfi:2011,Forbes:2011,Forbes:2012}.  The pairing functions in Eq.~(\ref{eq:BCS}) 
can be expressed as
\begin{equation}
\label{eq:pairfn}
\phi({\bf r}) = \sum_{n} \alpha_{n} e^{i {\bf k}_{\bf n} \cdot {\bf r}} + \tilde{\beta}(r) \,,
\end{equation}
which contains the variational parameters $\alpha_{n}$ for each momentum state up to some level $n$.  When setting $\alpha_n = 0$ for $|{\bf k}_{\bf n}| > k_F$ and $\tilde{\beta}(r)=0$ in Eq.~(\ref{eq:pairfn}), the BCS wave function in Eq.~(\ref{eq:BCS}) is equivalent to the Slater form in Eq.~(\ref{eq:Slater}).

Higher-momentum contributions are accounted for by the spherically symmetric beta function:
\begin{gather}
\tilde{\beta}(r) = \beta(r)+\beta(L-r)-2 \beta(L/2)~~~ \mbox{for}~~ r \le L/2\,, \nonumber  \\
~~~~~~ = 0 ~~~~~~~~~~~~~~~~~~~~~~~~~~~~~~~~~~~~ \mbox{for} ~~ r > L/2\,,
\nonumber  \\
\beta (r) = [ 1 + c b  r ]\ [ 1 - e^{ - d b r }] \frac{e^{ - b r }}{d b r}~,
\label{eq:beta}
\end{gather}
which contains the variational parameters $b$, $c$ and $d$.  This form of the beta function has been used for 3D calculations, and we have explicitly checked that it generalizes to 2D.  
Specifically, singular terms like $(1/r) \partial \tilde{\beta} / \partial r$ can cause large fluctuations in the local energy for small $r$, therefore $c$ is defined such that $\partial \tilde{\beta} / \partial r = 0$ at $r=0$.  For $b=0.5k_F$ and $d=5$~\cite{Gandolfi:2011}, we find $c \simeq 3.5$.

Jastrow correlation functions $f_{J}(r_{ij'})$ are introduced for each pair of interacting particles.  For example, a system with $N_\uparrow=3$ and $N_\downarrow=2$ has $3\times 2=6$ Jastrow terms resulting from the product over $i,j'$ in Eq.~(\ref{eq:BCS}).  By definition, $f(r_{ij'})$ is always positive and reaches 1 at the ``healing distance'', where it has zero 
derivative.  It's determined 
from the radial Schr\"odinger equation for $R(r)$.  This is equivalent to $u(r) / \sqrt{r}\,$, which follows from the solution to Eq.~(\ref{eq:swave}) where the scattering energy is adjusted to give a nodeless result.  
Collectively, the Jastrow terms reduce statistical errors.  In principle, the calculations in this work are independent of Jastrow correlations, though in practice the
2D DMC Jastrowless runs exhibited very large variance.

\begin{figure}[b]
\centering
\includegraphics[width=0.45\textwidth]{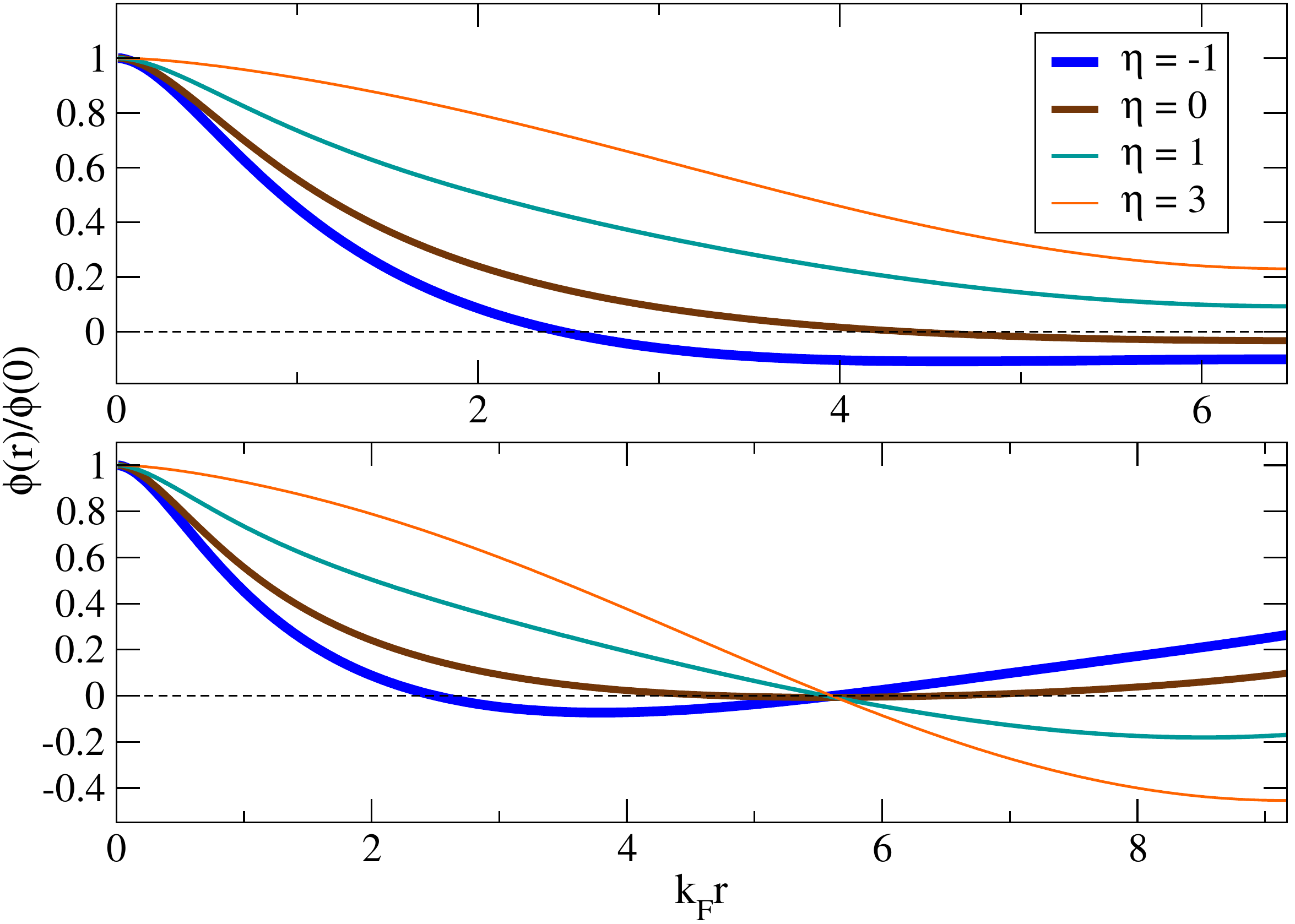}
\caption{ 
Variationally optimized pairing functions for various interaction strengths.  These are labelled by line thickness, i.e., the thickest line is for $\eta=-1$ and the thinnest 
line is for $\eta=3$.
The top panel plots $\phi({\bf r})$ through a path parallel to the box sides, corresponding to the (1,0) or equivalently the (0,1) direction.  A path along the (1,1) direction is taken for $\phi({\bf r})$ in the bottom panel plot.  The dashed lines are included to show node locations, where $\phi({\bf r})=0$. \label{fig:pair}}
\end{figure}

Parameters in Eq.~(\ref{eq:pairfn}) are optimized using Variational Monte Carlo (VMC) energy minimization (further discussed in section~\ref{sec:QMC}).  Unique sets of $\alpha_n$ were determined for each interaction strength, using automated optimization techniques~\cite{Sorella:2001}.  A selection of optimized pairing functions are plotted in Fig.~\ref{fig:pair}.  These two-body pairing orbitals contain information about the many-body system via the values arrived at for our $\alpha_n$ parameters.  They become increasingly localized about the origin as $\eta$ is reduced and pairs become more tightly bound.  The domain of $\phi({\bf r})$ in each panel is equal to the maximum distance along the respective path between particles, as discussed in the caption.  Nodes exist where $\phi({\bf r})$ crosses the r-axis (marked with a dashed line).  These occur in the top panel for $\eta=0$ and $-1$, where stronger attractions result in more localization.  For the longer path, shown in the bottom panel, a similar trend is seen.  Here we find that the stronger $\eta=-1$
case has an extra node. This and all other interaction strengths 
lead to a similar node farther out, which we have checked
is near the Jastrow-Slater node location.  For $\eta=6$ (not shown), we find the node slightly shifted to the left, even closer to the Slater node location.

\section{Quantum Monte Carlo}
\label{sec:QMC}
Quantum Monte Carlo methods allow for an accurate ground-state description of many-body systems from first principles, and are important for constraining theoretical models and guiding experiment~\cite{Foulkes:2001}.  Our techniques are variational in nature, which means that calculations give exact or upper bound estimates of energy expectation values.  One major complication of QMC for Fermi gases is the fermion-sign problem.  Fixed-node diffusion Monte Carlo circumvents the sign problem by restricting transitions across nodal surfaces.

Nodes are first optimized using Variational Monte Carlo, where the Metropolis algorithm is used to sample a trial wave function $\Psi_T({\bf R})$.  We calculate the variational estimate:
\begin{equation}
\label{eq:vmcener}
E_V = \frac{\int d{\bf R}\Psi^*_T({\bf R}) \hat{H} \Psi_T({\bf R})}{\int d{\bf R}| \Psi_T({\bf R})|^2}\,.
\end{equation} 
Our optimization requires many VMC iterations where $E_V$ is calculated repeatedly as parameters in $\Psi_T({\bf R})$ are adjusted.  The final parameters selected are those that give the lowest $E_V$.

After optimization, we acquire a set of equilibrated VMC configurations.  DMC is then used to project the ground state from the optimized trial wave function:
\begin{equation}
\label{eq:dmc1}
\Phi_0 = \lim_{\tau\rightarrow\infty}\Psi(\tau) \,,
\end{equation}
where
\begin{equation}
\label{eq:dmc2}
\Psi(\tau) = e^{-(\hat{H}-E_T)\tau}\Psi(0) \,,
\end{equation} 
and $\Psi(0)\equiv\Psi_T$.  The trial energy $E_T$ is a constant offset applied to the Hamiltonian and is important for controlling simulations.  DMC expectation values are determined by averaging over a set of equilibrated configurations.  For the energy, we calculate the mixed estimate:
\begin{equation}
\label{eq:dmcener}
H_M = \frac{\langle \Psi_T | \hat{H} | \Psi(\tau) \rangle}{\langle \Psi_T | \Psi(\tau) \rangle} \,.
\end{equation}

The projection in Eq.~(\ref{eq:dmc2}) is accomplished by using Green's functions to propagate the wave function in time steps of size $\Delta\tau$.  In order to evaluate these Green's functions, the Trotter-Suzuki approximation is used, which requires that $\Delta\tau$ be very small.  We reduce the DMC time step $\Delta\tau$ until any errors introduced by the approximation are much smaller than statistical error.  As could be expected, the maximum acceptable time-step size decreases as the energy scale increases.  To account for this, we reduced $\Delta\tau$ as we probed deeper into the BEC side of the crossover.

Computational runtimes are roughly proportional to $N^3$, thereby limiting the maximum 
possible particle number.  We've done tests in 2D for system sizes up to $N=58$ and found good agreement with $N=26$ results.  The non-interacting energy per particle for the $N=26$ closed shell is quite close to $E_{\rm FG}$ (see inset of Fig.~\ref{fig:finite}).  As in other
systems with strong pairing~\cite{Forbes:2011}, such free-particle shell effects get ``washed out''. Finite-size corrections are therefore small, as discussed below.

\section{Equation of State}
\label{sec:results}
\begin{figure}[t]
\centering
\includegraphics[width=0.45\textwidth]{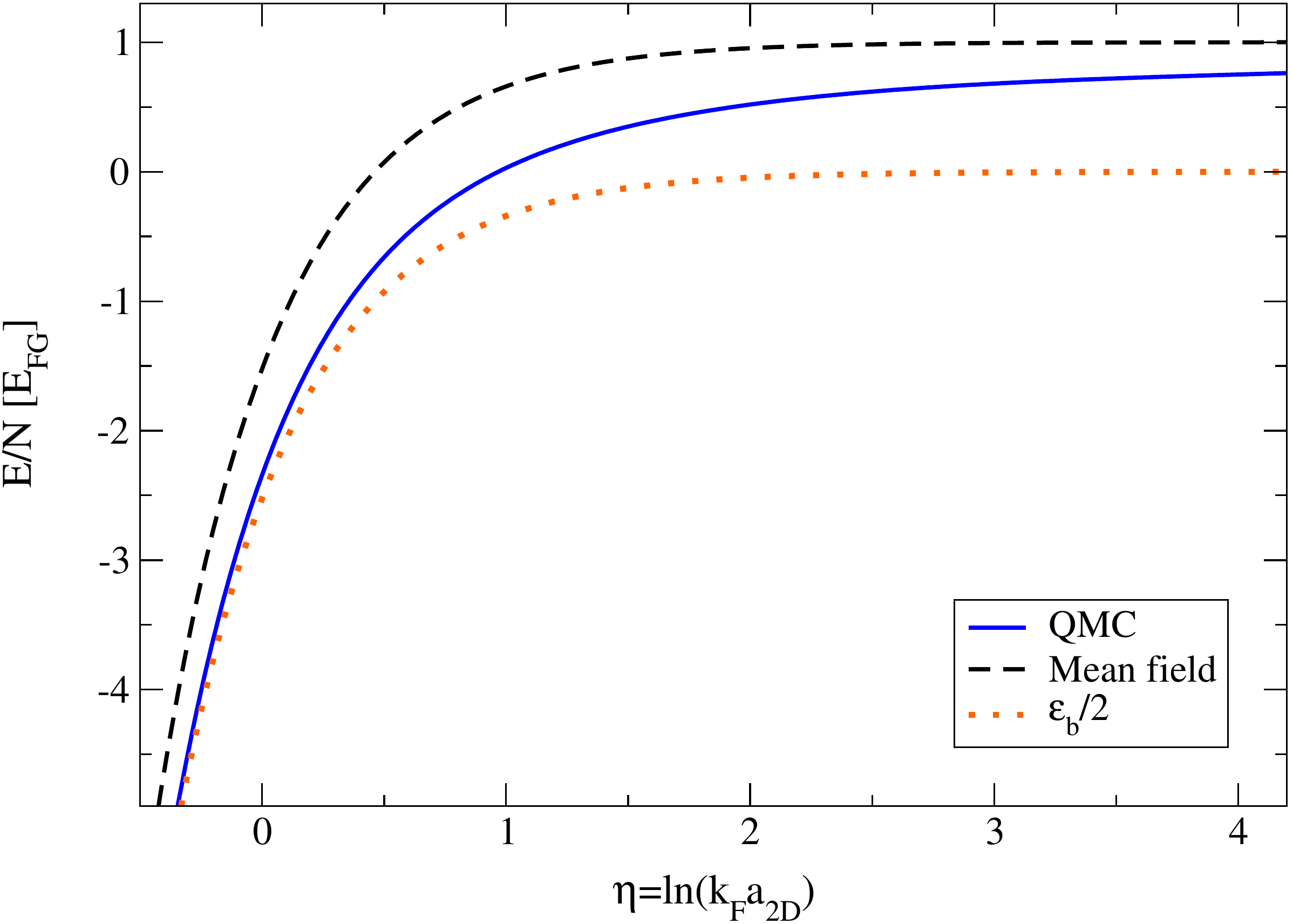}
\caption{Total energy per particle of the 2D strongly interacting Fermi gas.  
Thermodynamic-limit-extrapolated QMC results are plotted with a solid line (blue).  At this energy resolution, all QMC results are in good agreement.  The mean-field result is shown as a dashed line (black) for comparison.  This is expected to be reasonably accurate for $|\eta| \gg 1$.  Half of the binding energy per particle is plotted with a dotted line (orange).  There is logarithmically small binding energy in the BCS limit and $|\epsilon_b|$ becomes very large in the BEC limit.  \label{fig:eos1}}
\end{figure}

Figure~\ref{fig:eos1} shows the ground-state energy per particle in the BEC-BCS crossover.  The mean-field description gives an energy per particle equal to $E_{\rm BCS} = E_{\rm FG} + \epsilon_b/2$~\cite{Randeria:1989}.  Mean-field theory is expected to be accurate for weakly paired systems in the BCS limit (when $E/N \rightarrow E_{\rm FG}$).  The BEC limit of tightly bound pairs is also expected to be reasonably well described by mean field.  On that side of the crossover, the energy scale grows rapidly by many orders of magnitude due to large binding energies.  The QMC results become increasingly similar to mean-field predictions in each limit.  Differences between our calculations and previous QMC results only become clear when subtracting the two-body binding energy contribution.  We will compare to these studies in detail later in this section.

\begin{figure}[t]
\centering
\includegraphics[width=0.45\textwidth]{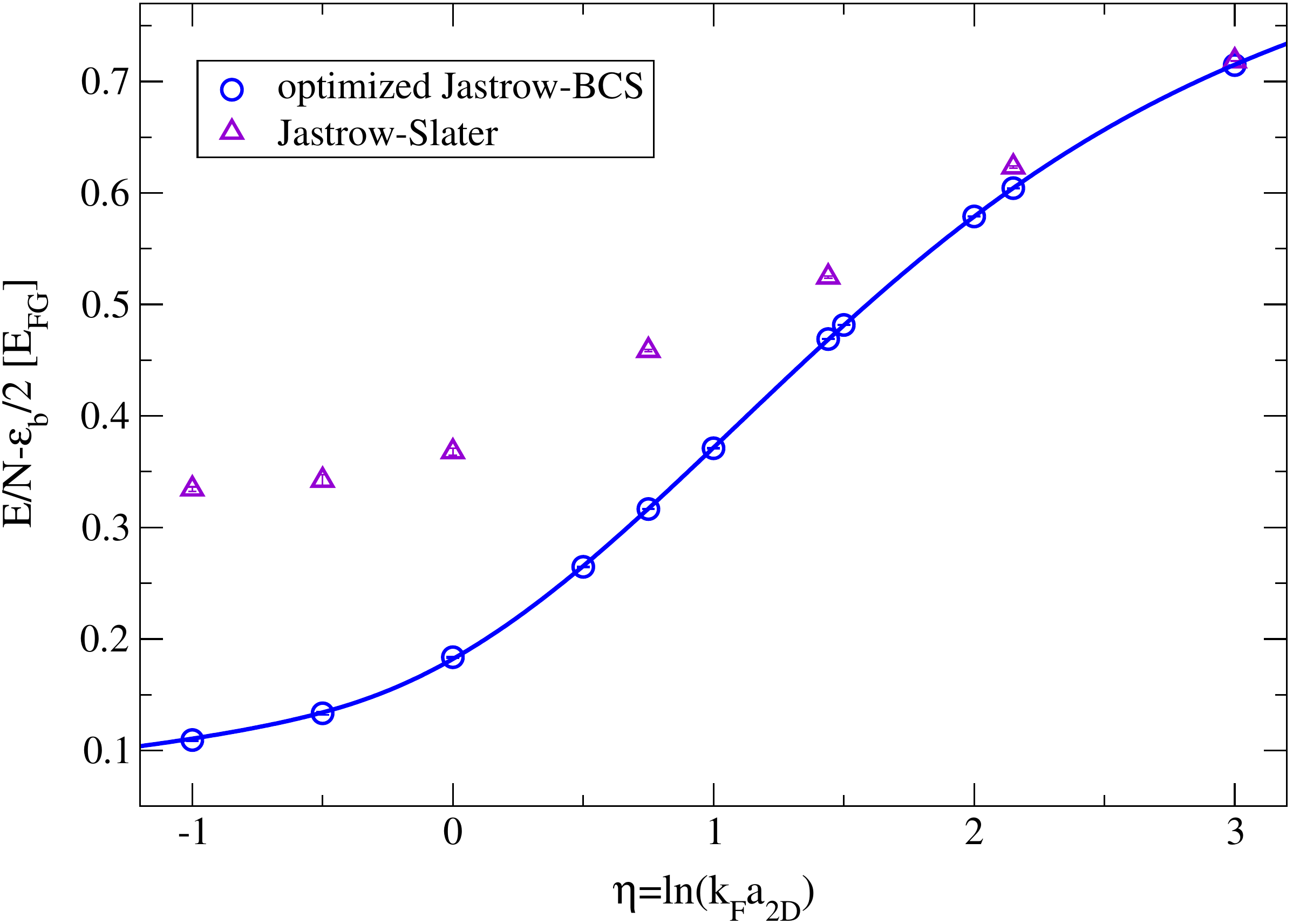}
\caption{Our DMC energy results for the strongly interacting Fermi gas in the BEC-BCS crossover.  The binding energy per particle has been subtracted and the units are $E_{\rm FG} = \hbar^2 k^2_F / 4m$.  Our DMC calculations for the Jastrow-BCS and Jastrow-Slater wave functions are shown with circles (blue) and triangles (purple), respectively.  
With decreasing $\eta$,
we see increasingly significant improvements over the Slater energy results
by using the optimized BCS wave function.
 \label{fig:eos2}}
\end{figure}

The question arises of which wave function to use. We have introduced two many-body wave functions in Sec.~\ref{sec:interacting}.  We test the Jastrow-Slater wave function $\Phi_{\rm S}$, Eq.~(\ref{eq:Slater}), for the strongly interacting gas to quantify its shortcomings.  This wave function has a long history of describing weak pairing in quantum gases.  It can be seen from previous works on the 2D Fermi gas~\cite{Bertaina:2011,Shi:2015} that $\Phi_{\rm S}$ describes the weakly paired regime well on the BCS side of the crossover.  Because it only accounts for pairing through the Jastrow component, $\Phi_{\rm S}$ is not expected to provide reliable ground-state energies when the coupling is sufficiently strong.  Also, this form has limited freedom to optimize the many-body wave-function nodes.  
The Jastrow-BCS wave function $\Phi_{\rm BCS}$, Eq~(\ref{eq:BCS}), contains parameters which are variationally optimized for each $\eta$ independently.  It is possible to choose these undetermined $\alpha$ parameters such that $\Phi_{\rm BCS}$ has the same nodes as $\Phi_{\rm S}$.  This is generally the set of $\alpha$'s used as the starting point for optimization.

Our DMC energy results are shown in Fig.~\ref{fig:eos2}, where half of the two-body binding energy $\epsilon_b$ has been subtracted.  The energy per particle is plotted as a function of interaction strength $\eta = \ln(k_Fa_{\rm 2D})$ in units of $E_{\rm FG}$.  Error bars represent statistical uncertainty, which is much smaller than the symbol used to mark each point.  The triangles are calculated using the Jastrow-Slater wave function~\cite{Note:1}; for $\eta<2$ we see significant improvement by optimizing the pairing function.  In this plot, the mean-field result would correspond to a horizontal line at 1.  In the BEC limit of tightly bound pairs, however, it appears that $E/N \rightarrow \epsilon_b/2$.  Although the QMC results seem to suggest a trend away from the BEC mean field prediction, the binding energy becomes very large in this limit such that $|\epsilon_b| \gg E_{\rm FG}$ and $E_{\rm BCS} \approx \epsilon_b/2$.

\begin{figure}[t]
\centering
\includegraphics[width=0.45\textwidth]{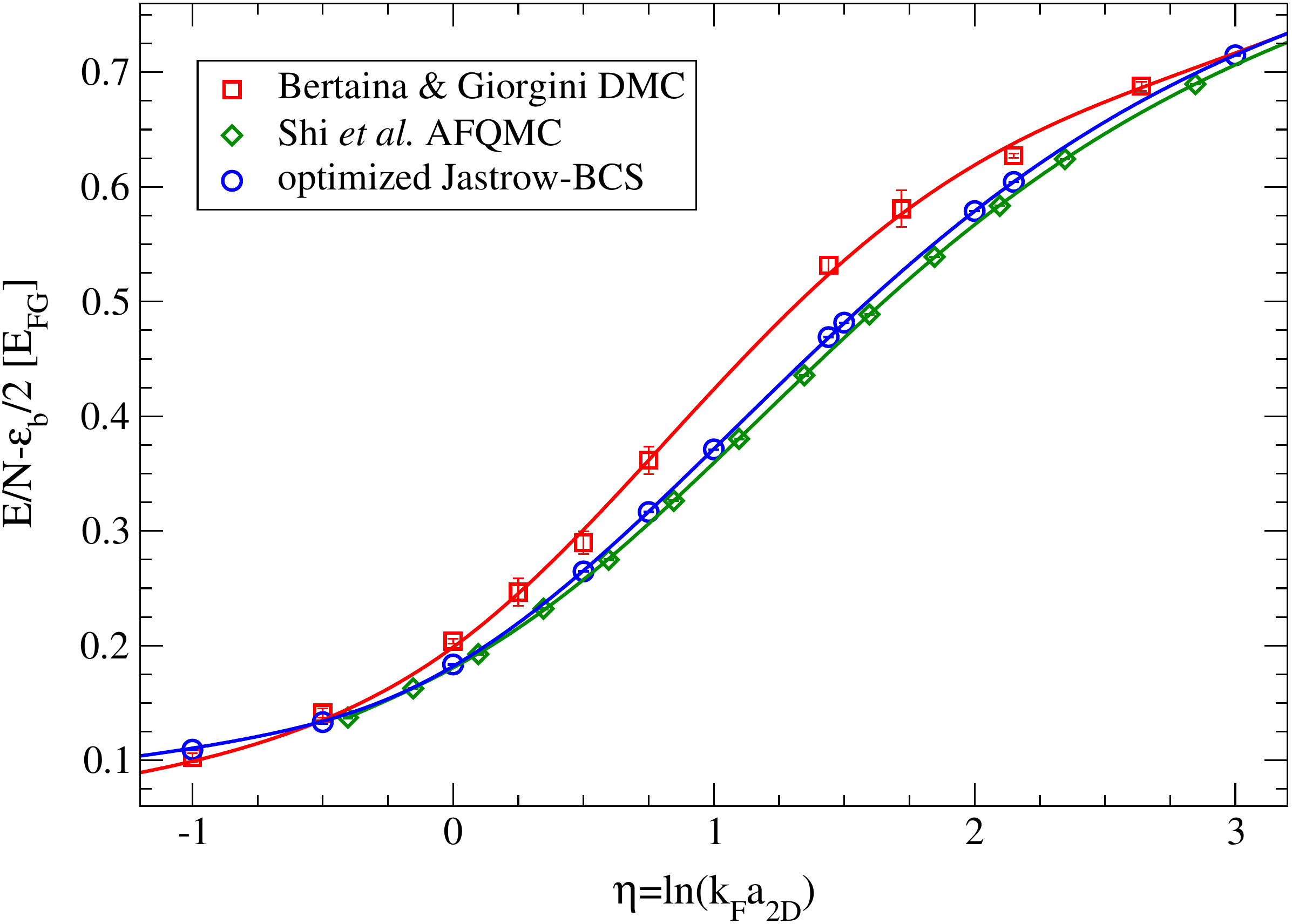}
\caption{The energy-per-particle results for various {\it ab initio} QMC methods are compared.  Our optimized Jastrow-BCS wave function results are shown with circles (blue).  These closely follow the AFQMC energy results of Shi {\it et al}~\cite{Shi:2015}, plotted as diamonds (green).  Previous DMC energy results of Bertaina \& Giorgini~\cite{Bertaina:2011} 
are plotted as squares (red); since DMC is variational, the new results show a notable improvement. Note that the error bars, which are provided for all data, are comparable to the symbol size (or larger) only for previous DMC results.  \label{fig:eos3}}
\end{figure}

We compare to previous DMC~\cite{Bertaina:2011} and AFQMC~\cite{Shi:2015} results for the strongly interacting 2D gas in Fig.~\ref{fig:eos3}.  The previous DMC results were calculated using a square well potential to model two-body interactions, and wave functions with only one variational parameter (at most); a Jastrow-Slater wave function was used for $\eta>1$, and a BCS-pairing wave function was used for $\eta<1$.  In our calculations, we use a continuous potential, Eq.~(\ref{eq:potential}), that decays smoothly.  More importantly, we optimize 10 variational $\alpha$ parameters in the pairing function in Eq.~(\ref{eq:BCS}), for each calculation.  We see a notable improvement from previous DMC ground-state energy results in the crossover region.  The DMC method is variational, so our lower energy results are closer to the true ground-state energy.  We directly compare results for $\eta=0.75$, $\eta=1.44$, and $\eta=2.15$~\cite{Note:1}, and find improvements of ~12\%, 12\%, and 6\%, respectively, 
on the scale of Fig.~\ref{fig:eos3}.  The AFQMC method used by Shi {\it et al} is 
in principle exact, in the sense that it does not suffer from the sign problem for $N_{\uparrow} = N_{\downarrow}$ configurations.  Therefore, limitations of the fixed-node approximation can be
seen by comparing our DMC results with AFQMC.  The error introduced by this approximation is quite small: differences are 0.014~$E_{\rm FG}$ at most.  For comparison, in 3D the AFQMC energy per particle for the unitary Fermi gas is 0.372(5)~$E_{\rm FG}$~\cite{Carlson:2011}; this can be compared to a detailed DMC study which finds an energy of 
$\sim$0.3897(4)~$E_{\rm FG}$~\cite{Forbes:2012}. While this result was for one interaction strength in the middle of the 3D crossover, we find a similar difference for a range of $\eta$ on the BCS side of the 2D crossover.  For $\eta < 1$, we find increasingly smaller differences between DMC and AFQMC as we approach the BEC limit.

All energy results shown up to this point (in Figs.~\ref{fig:eos2} and \ref{fig:eos3}) are for $N=26$; such a system size is 
well suited to simulate the $N \rightarrow \infty\,$ 2D system in the region 
studied here.  Specifically, after finite-size corrections are applied, the $N=26$ AFQMC energies calculated by Shi {\it et al} differ from their $N \rightarrow \infty$ energies by 0.003~$E_{\rm FG}$ or less for interaction strengths $\eta<1$.  For $\eta>1$ the maximum difference from the AFQMC TL value is 0.01~$E_{\rm FG}$ at $\eta \approx 2$.  Finite-size effects (first touched upon 
in section~\ref{sec:free}) are addressed by applying a TL correction to QMC energy calculations.  Specifically, the finite-size effects for the Slater wave function are illustrated in Fig.~\ref{fig:finite}.  Here we see reasonably small variation from the TL for $N=26$: for very large $\eta$ (say $\eta>4$) such a correction would essentially all come
from the kinetic energy and would be $\sim$0.041~$E_{\rm FG}$; this matches the correction
 used by Bertaina \& Giorgini~\cite{Bertaina:2011} (from Fermi liquid theory) at $\eta \gtrsim 1$.  For $\eta<4$, when using the BCS wave function, the finite-size effects are greatly reduced since pairing is taken into account. 
 One way to estimate their magnitude is to solve the 
 mean-field BCS problem for both infinite and finite-size systems and compare the two,
 as was done in Ref.~\cite{Gezerlis:2008} for neutrons. Shi {\it et al} have done this for
 the 2D Fermi gas, with the results being indeed quite small in the crossover: 
 0.001~$E_{\rm FG}$ at $\eta \approx 1$ and 0.01~$E_{\rm FG}$ at 
 $\eta \approx 2$
 ~\cite{Shi:2015,Shi:2015b}. It is easy to see that no finite-size 
 correction is necessary in the deep BEC regime (or near it).

\begin{figure}[t]
\centering
\includegraphics[width=0.45\textwidth]{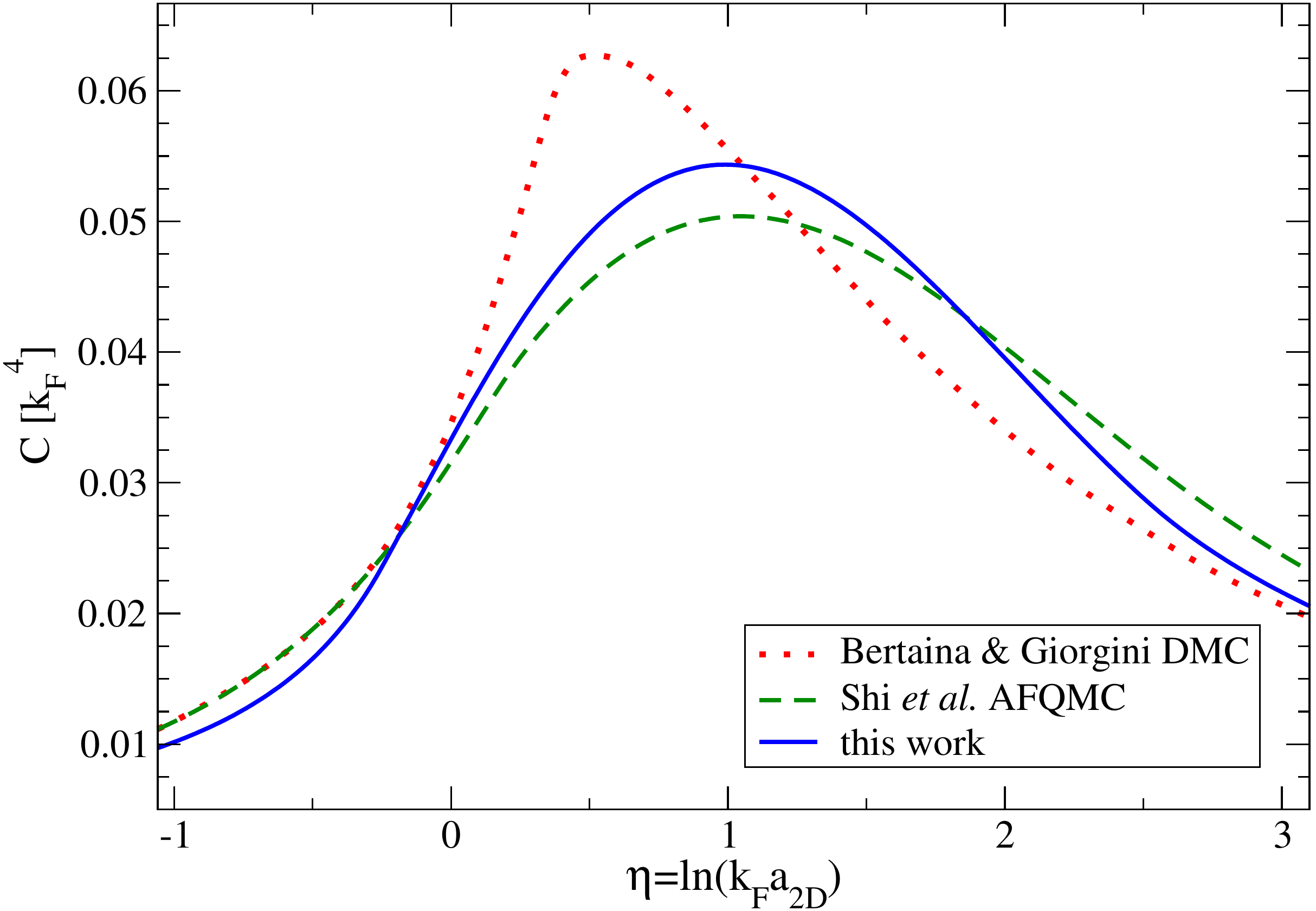}
\caption{The contact parameter as determined by the EOS derivative.  The result from our DMC method is plotted as the solid line (blue) and we compare to other QMC results.  We show previous DMC results~\cite{Bertaina:2011} with a dotted line (red), and AFQMC
results~\cite{Shi:2015} with a dashed line (green).  Qualitatively, the contact reaches a maximum on the BCS side of the strong coupling regime and decays more rapidly on the BEC side of the crossover.} \label{fig:contact}
\end{figure}

Our TL equation of state is parametrized using similar methods as in Refs.~\cite{Bertaina:2011,Shi:2015}.  We fit to a 7th-order polynomial for the crossover region: $f(\eta) = \sum^{7}_{i=0}c_i\eta^i$, a dimer form in the BEC regime (see Ref.~\cite{Levinsen:2014}), and an expansion in $1/\eta$ in the BCS regime (see Ref.~\cite{Engelbrecht:1992}).  For $\eta<-0.25$ we fit to:
\begin{equation}
\label{eq:BECform}
f^{\rm BEC}(\eta) = \frac{1}{2x} \bigg[1-\frac{\ln(x)+d}{x}+\frac{\sum_{i=0}^2 c_i [\ln(x)]^i}{x^2}\bigg]\,,
\end{equation}
where $x = \ln[4\pi/(k_F a_d)^2] \approx 3.703 - 2\eta$ (for the dimer scattering length $a_d \approx 0.557a_{\rm 2D}$) and $d = \ln\pi + 2\gamma + 0.5$.  For $\eta>2.5$ we fit to:
\begin{equation}
\label{eq:BCSform}
f^{\rm BCS} (\eta) = 1 - \frac{1}{\eta} + \sum_{i=2}^{4}\frac{c^i}{\eta^i}\,.
\end{equation}
Values of $c_i$ in Eq.~(\ref{eq:BECform}) and Eq.~(\ref{eq:BCSform}) are determined using continuity conditions for $f$, $\partial f/\partial \eta$, and $\partial^2 f/\partial \eta^2$ at the matching points.  This method is also used to produce the lines of best fit seen in Fig.~\ref{fig:eos3} for Bertaina \& Giorgini and Shi {\it et al}.  We pick our matching points to optimize the overall fit and include as much of the crossover polynomial function as possible.  We ensure that our matching point for Eq.~(\ref{eq:BECform}) is selected on the BEC side of the crossover.

The contact parameter is universal for dilute Fermi gases~\cite{Tan:2008}, and can be related to thermodynamic quantities including the pressure and chemical potential.  It describes the momentum distribution $n(k)$ behaviour at large $k$ and therefore encodes information about short-range physics.  In 3D the contact parameter's magnitude steadily
declines as one goes from the BEC to the BCS regime~\cite{Hoinka:2013}. Qualitatively,
something similar happens in 2D. In order to bring out more detailed effects, we subtract 
the two-body binding energy contribution and thereby relate our EOS to the 
(``many-body'') contact density as follows~\cite{Werner:2012,Bertaina:2011,Shi:2015}: 
\begin{equation}
\frac{C}{k^4_F} = \frac{1}{4} \frac{d[(E/N)/E_{\rm FG}]}{d\eta}
- \frac{1}{4} \frac{d[(\epsilon_b/2)/E_{\rm FG}]}{d\eta}
\end{equation}
In Fig.~\ref{fig:contact}, we compare our results with other QMC approaches.  We find a maximum value of $C/k^4_F\approx 0.054$ at $\eta\approx 1$.  Our curve is slightly different than Shi {\it et al} who find a maximum of  $C/k^4_F\approx 0.05$ at roughly the same $\eta$.  Bertaina \& Giorgini find a higher and significantly more narrow peak at $\eta \approx 0.5$.

\section{Pairing gap}
\label{sec:gap}
The single-particle excitation-spectrum pairing gap can be calculated by comparing the total energy of configurations with an unpaired particle to that of fully paired-up systems (i.e., from the odd-even energy staggering)~\cite{Carlson:2003}.  For the 2D gas, the only previous \textit{ab initio} calculation of the pairing gap is the one by Bertaina \& Giorgini~\cite{Bertaina:2011}.  Because we have variationally improved the EOS, it is justified to re-calculate the pairing gap using our new many-body wave functions.  
For the case when $N_\uparrow\neq N_\downarrow$, the BCS wave function in Eq.~(\ref{eq:BCS}) is no longer valid.  Adding one additional particle to an even system such that we have $M=(N-1)/2$ pairs, the wave function can be written as:
\begin{equation}
\Phi_{\rm BCS} = {{\cal A}}[\phi({\bf r}_{11'}) \phi({\bf r}_{22'}) ... \phi({\bf r}_{M M'}) \psi_{{\bf k}_u}({\bf r})]\,,
\end{equation}
where the unpaired particle at position ${\bf r}$ and momentum ${\bf k}_u$ is placed into a plane-wave eigenstate.  The extra particle's momentum must be treated as a variational parameter.  For $N=27$ systems (with one unpaired particle), we found the optimal $|{\bf k}_u|$, in units of $2\pi /L$, to be 0 for $\eta \lesssim -1$ and $\sqrt{5}$ for $\eta \gtrsim 3$.  In the crossover region we see a smooth transition as more (or fewer) momentum levels become available to the extra particle.  Note that for our $N=26$ system, the maximum k-state occupied in the Slater case is $k_{\rm max} = (2\pi/L) \sqrt{5}$.  Our results are consistent with variational calculations for the 3D Fermi gas, which find $|{\bf k}_u| = 0$ in the BEC limit and $|{\bf k}_u| = k_{\rm max}$ in the BCS limit~\cite{Chang:2004}.

\begin{figure}[b]
\centering
\includegraphics[width=0.45\textwidth]{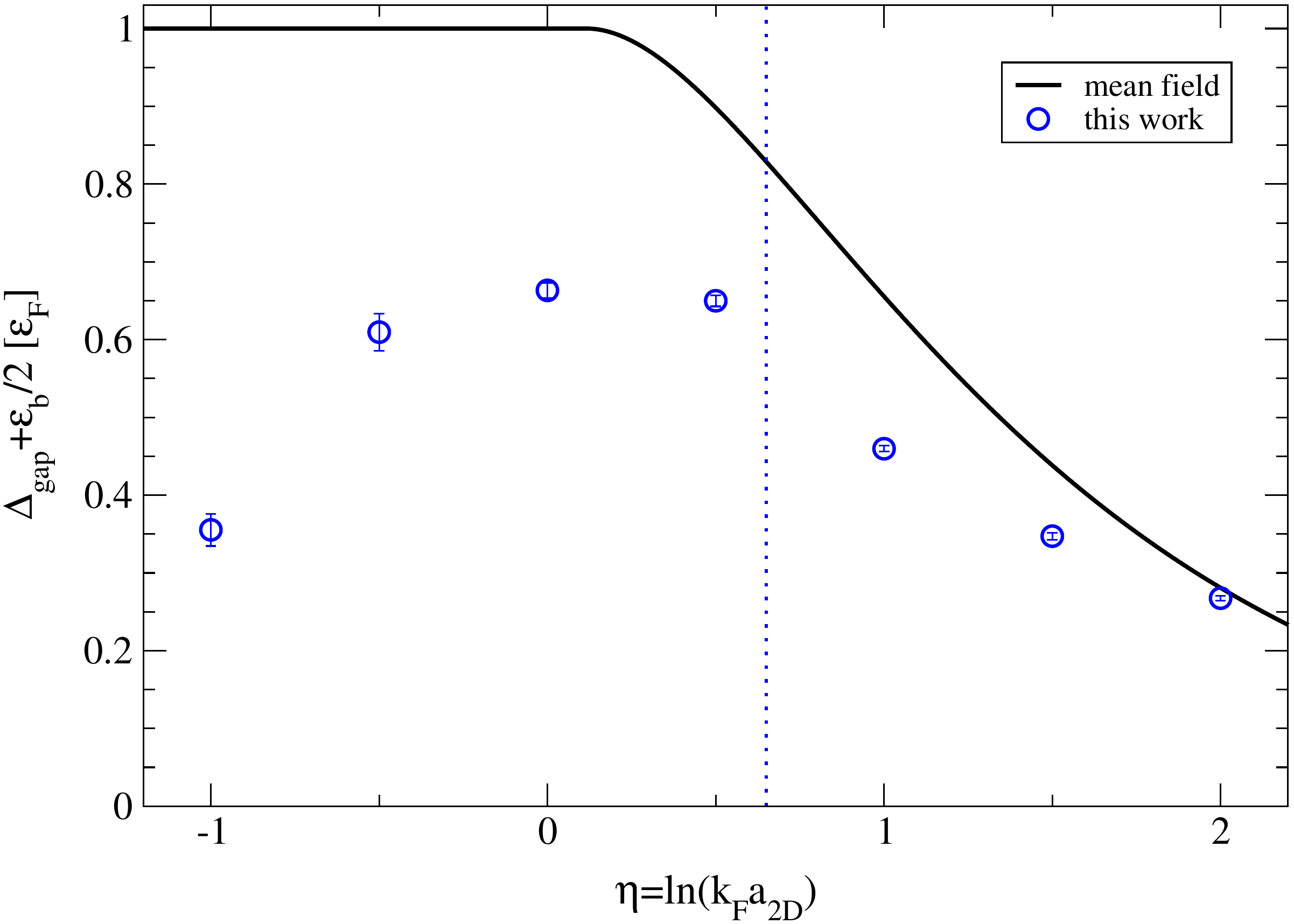}
\caption{The pairing gap in the strong coupling crossover.  The binding energy per particle has been subtracted and the units are $\epsilon_F = \hbar^2k^2_F / 2m=2 \, E_{\rm FG}$.  Our results are shown with circular symbols (blue) and the mean-field prediction is shown with a solid line (black).  Our calculated chemical potential $\mu$ becomes zero for the interaction strength $\eta \approx 0.65$.  This location is marked with a dotted line (blue).  Mean-field theory predicts that $\mu = 0$ at $\eta \approx 0.12$.} \label{fig:gap}
\end{figure}

To determine the pairing gap, we use the odd-even energy staggering:
\begin{equation}
\Delta_{\rm gap} = E(N+1) - \frac{1}{2}[E(N)+E(N+2)]\,,
\end{equation}
where $N$ is an even number.  Accurate calculations are possible for cold atomic gases because $\Delta_{\rm gap}$ is a large fraction of the total energy.  Each of our points in Fig.~\ref{fig:gap} is the result of 3 separate DMC simulations.  We find a broad peak about $\eta=0$ at $\Delta_{\rm gap} + \epsilon_b/2 \approx 0.65 \, \epsilon_F$.  The mean-field description identifies $\Delta_{\rm gap}$ as the minimum Bogoliubov quasiparticle energy, giving $\Delta_{\rm gap} = \sqrt{2\epsilon_F|\epsilon_b|}$ for positive $\mu$ when $\epsilon_F > |\epsilon_b|/2$ and $\Delta_{\rm gap} = \epsilon_F + |\epsilon_b|/2$ for negative $\mu$ when $\epsilon_F < |\epsilon_b|/2$~\cite{Miyake:1983,Randeria:1989}.  The mean-field chemical potential is given by $\mu = \epsilon_F + \epsilon_b/2$, and the transition between gap equations at $\mu=0$ corresponds to $\eta \approx 0.12$.  Evaluating the derivative $\partial E / \partial N$ we find that for our strongly coupled theory the chemical potential changes
sign at $\eta \approx 0.65$.

Considering the energy difference between mean field and QMC in Fig.~\ref{fig:eos1}, it's not expected that the mean field will give accurate pairing gaps in the crossover: our gaps 
in the strongly coupled regime of the crossover are smaller than the mean-field values.  This was first found by Bertaina \& Giorgini~\cite{Bertaina:2011}, but our detailed values are different due to improved optimization.  Our results follow a smoother trend
and have significantly smaller error bars. Note that as we go deep into the BCS regime,
one expects from the theory of Gorkov and Melik-Barkhudarov that the pairing gap
should be suppressed by a factor of $e$ with respect to the BCS 
value~\cite{Petrov:2003}. (This is to be compared
with the factor of $(4e)^{1/3}$ that appears in the 3D case~\cite{Gorkov:1961,Gezerlis:2008}).
Our (finite-size uncorrected) results do not exhibit such a suppression.
Of course, it is very difficult to extract a pairing gap from DMC simulations for $\eta \gg 1$,
where the gap is small.

\section{Summary and outlook}

In summary, we have performed {\it ab initio} calculations for the strongly interacting two-component Fermi gas in the BEC-BCS crossover.  We compared our energy results 
to calculations from other QMC methods.  We see significant improvements from previous DMC results in the strongly interacting crossover regime.  When comparing to AFQMC results, our EOS is very similar, as a result of variationally minimizing the effect of the fixed-node approximation.  All QMC calculations are in good agreement in the BCS side of the crossover for $\eta\gtrsim 3$.  In this regime, the Jastrow-Slater wave function, Eq.~(\ref{eq:Slater}), can provide a good description of the weak many-body pairing.  For $\eta\lesssim 3$, we showed the Jastrow-Slater wave function is inadequate (see Fig.~\ref{fig:eos2}), and the Jastrow-BCS pairing wave function, Eq.~(\ref{eq:BCS}), provides a considerably better description.

Using our EOS, we also determined the contact parameter.  This looks similar to the AFQMC result and peaks at roughly the same interaction strength $\eta\approx 1$.  We also determined the pairing gap in the crossover regime; these are currently the most dependable predictions available for the 2D Fermi gas.  After subtracting the binding energy per particle, we find the gap to be a significant fraction of the Fermi energy in the strongly coupled crossover regime.  Approaching the BCS limit, our pairing gap drops in magnitude and as a result appears to 
tend towards the mean-field prediction.

There are many research opportunities that naturally follow from the present work.  
The pairing evolution between dimensions could be studied by probing the transition of the Fermi gas between 2D and 3D.  This has already been achieved in experiment~\cite{Dyke:2011,Sommer:2012}.  Our study could be extended to the quasi-2D regime by introducing a finite box-length 
or an external trap in the third dimension. Another possible research avenue involves a two-component Fermi gas in mixed dimensions, where one species is confined to 2D and the other is free to move in 3D space~\cite{Nishida:2008}. Two-dimensional polarized gases would also
be amenable to a study along the lines of the present work. Furthermore, the 2D Fermi gas could  be studied under the influence of a periodic external potential.  This model could describe a two-component gas in an optical lattice~\cite{Pilati:2014,Carlson:2014}, and has recently been applied to the problem of 
neutron matter interacting with a lattice of ions in a neutron star~\cite{Buraczynski:2015}.

\

\begin{acknowledgments}
The authors would like to thank Hao Shi and Shiwei Zhang for useful discussions and for
sharing the results of their calculations. This work was supported in part 
by the Natural Sciences and Engineering Research Council (NSERC) of Canada, the 
Canada Foundation for Innovation (CFI), the US Department of Energy,
Office of Nuclear Physics, under Contract DE-AC52-06NA25396, and the LANL LDRD program. 
Computational resources were provided by SHARCNET, NERSC, and Los Alamos Open Supercomputing.
\end{acknowledgments}


\begin{thebibliography}{99}
\bibitem{Note:1}
For the Jastrow-Slater calculations as well as when comparing Jastrow-BCS results
to Bertaina \& Giorgini we tune $V(r)$, Eq.~(\ref{eq:potential}), to match their effective range.
Explicitly, for $\eta=0.75$, 1.44, and 2.15, the values are $k_F r_e \approx 0.0068, 0.0071,$ and $0.0075$ respectively. We have checked that there is no noteworthy effect 
arising from this minor variation.
\end{thebibliography}
\end{document}